\PassOptionsToPackage{hyphens}{url}
\documentclass[10pt,conference]{IEEEtran}

\usepackage{cite}
\usepackage{amsmath,amssymb,amsfonts}
\usepackage{booktabs}
\usepackage{graphicx}
\usepackage{xspace}
\usepackage{xcolor}
\usepackage{tikz}
\usetikzlibrary{arrows.meta,positioning,fit}
\usepackage{pifont}                 
\usepackage[hidelinks]{hyperref}
\usepackage[capitalize,nameinlink]{cleveref}
\usepackage{relsize}
\usepackage[numbers,sort&compress]{natbib}
\usepackage[nogroupskip,nonumberlist,nopostdot]{glossaries}
\newacronym{eip}{EIP}{Ethereum Improvement Proposal}

\newacronym{ml}{ML}{Machine Learning}

\newacronym{evm}{EVM}{Ethereum Virtual Machine}
\newcommand{\evm}{\gls*{evm}\xspace}
\newacronym{defi}{DeFi}{Decentralized Finance}
\newcommand{\defi}{\gls*{defi}\xspace}
\newacronym[plural={LLMs}]{llm}{LLM}{Large Language Model}
\newcommand{\llm}{\gls*{llm}\xspace}
\newcommand{\llms}{\glspl*{llm}\xspace}
\newacronym[plural={APIs}]{api}{API}{Application Programming Interface}
\newcommand{\api}{\gls*{api}\xspace}

\newacronym{scwe}{SCWE}{OWASP Smart Contract Weakness Enumeration}
\newcommand{\scwe}{\gls*{scwe}\xspace}
\newacronym{bsc}{BSC}{BNB Smart Chain}
\newcommand{\bsc}{\gls*{bsc}\xspace}
\newacronym{claude}{Claude}{Claude~Opus~5.5}
\newcommand{\claude}{\gls*{claude}\xspace}
\newacronym{ci}{CI}{95\%~Confidence Interval}
\newcommand{\ci}{\gls*{ci}\xspace}

\DeclareRobustCommand*{\circnum}[1]{%
  \tikz[baseline=(char.base)]{
    \node[
      circle,
      draw=black,
      fill=black,
      text=white,
      line width=0.4pt,
      inner sep=0.2pt,
      outer sep=0pt,
      minimum size=0pt,   
      text width=         
    ] (char) {#1};
  }%
}

\newcommand{\code}[1]{{\smaller\texttt{#1}}\xspace}
\newcommand{\class}[1]{{\smaller\urlstyle{tt}\nolinkurl{#1}}\xspace}

\definecolor{darkred}{rgb}{0.75,0,0}

\newcommand{\kasolved}[1]{\textcolor{orange}{$\blacktriangleright$KA:\ding{51}?$\blacktriangleleft$}}

\newcommand{\txbench}{\textsc{TxBench}\xspace}

\newcommand{\accValue}{868.2~million\xspace}
\newcommand{\aucCallsDep}{0.99\xspace}

\newcommand{\blocksecFirstDate}{\mbox{2023-01-03}\xspace}
\newcommand{\blocksecLastDate}{\mbox{2026-07-24}\xspace}

\newcommand{\ciReviewA}{92.5--99.3\%\xspace}
\newcommand{\ciReviewB}{96.3--100\%\xspace}
\newcommand{\eulerLossBlockSec}{200~million\xspace}
\newcommand{\eulerRegValue}{8.7~million\xspace}

\newcommand{\eulerTxs}{6\xspace}

\newcommand{\eulerValue}{198.1~million\xspace}

\newcommand{\kappaCI}{0.55--0.95\xspace}
\newcommand{\kappaReview}{0.75\xspace}

\newcommand{\nAccExploit}{846\xspace}

\newcommand{\nAccKey}{166\xspace}

\newcommand{\nAccPreTwentyThree}{77\xspace}

\newcommand{\nAccUnvouched}{210\xspace}
\newcommand{\nAccepted}{1,012\xspace}

\newcommand{\nBlockSecIncidents}{308\xspace}

\newcommand{\nBlockSecOnly}{780\xspace}

\newcommand{\nBoth}{197\xspace}
\newcommand{\nCandidates}{316,235\xspace}
\newcommand{\nCited}{1,850\xspace}
\newcommand{\nDisagree}{five\xspace}

\newcommand{\nFailCheck}{123\xspace}

\newcommand{\nFailLoss}{62\xspace}

\newcommand{\nFullTrace}{1,797\xspace}

\newcommand{\nJudged}{1,797\xspace}
\newcommand{\nKeyPlain}{121\xspace}
\newcommand{\nLabeled}{14,993\xspace}
\newcommand{\nNeg}{2,024\xspace}

\newcommand{\nNegBehavioral}{826\xspace}

\newcommand{\nNegDeploy}{372\xspace}
\newcommand{\nNegFinancial}{496\xspace}

\newcommand{\nNegOrdinary}{330\xspace}

\newcommand{\nNegTraced}{14,994\xspace}
\newcommand{\nNonexistent}{5\xspace}

\newcommand{\nOverlap}{40\xspace}
\newcommand{\nPassChecks}{1,174\xspace}
\newcommand{\nRegistryOnly}{873\xspace}
\newcommand{\nRegistryProtocols}{48\xspace}
\newcommand{\nRegistryRows}{300\xspace}
\newcommand{\nReverted}{5\xspace}

\newcommand{\nReviewA}{100\xspace}
\newcommand{\nReviewAConf}{98\xspace}
\newcommand{\nReviewARejected}{2\xspace}
\newcommand{\nReviewB}{100\xspace}
\newcommand{\nReviewBConf}{100\xspace}
\newcommand{\nReviewStrata}{18\xspace}

\newcommand{\nScweId}{1,166\xspace}
\newcommand{\nScweNone}{228\xspace}

\newcommand{\nScweRevSet}{26\xspace}
\newcommand{\nScweSheet}{30\xspace}

\newcommand{\nSecondary}{162\xspace}

\newcommand{\nSucceeded}{1,840\xspace}
\newcommand{\nTakeValue}{1,297\xspace}
\newcommand{\nTarget}{2,024\xspace}

\newcommand{\nTierOne}{7,715\xspace}

\newcommand{\nTopIncident}{153\xspace}

\newcommand{\nTraced}{1,845\xspace}

\newcommand{\nWindows}{8,091\xspace}

\newcommand{\nZsAsked}{193\xspace}
\newcommand{\nZsAttacks}{91\xspace}
\newcommand{\nZsExploits}{77\xspace}
\newcommand{\nZsKey}{14\xspace}
\newcommand{\nZsNegatives}{102\xspace}
\newcommand{\nZsNotAsked}{7\xspace}

\newcommand{\pAccCreation}{18.4\%\xspace}
\newcommand{\pAccFresh}{43.9\%\xspace}
\newcommand{\pAccGas}{54.5\%\xspace}

\newcommand{\pAccUnvouched}{20.8\%\xspace}
\newcommand{\pAgree}{87.5\%\xspace}

\newcommand{\pBlockGas}{1\%\xspace}

\newcommand{\pBoth}{10.6\%\xspace}

\newcommand{\pNegDeployNoCall}{77.2\%\xspace}

\newcommand{\pNegGas}{13.3\%\xspace}
\newcommand{\pReviewA}{97.7\%\xspace}
\newcommand{\pReviewB}{100\%\xspace}

\newcommand{\pTakeValue}{72.2\%\xspace}

\newcommand{\pTopIncident}{15.1\%\xspace}
\newcommand{\pZsAlways}{47\%\xspace}

\newcommand{\sizeGB}{5.4\xspace}
\newcommand{\topIncidentName}{BtcTurk\xspace}
\newcommand{\zsAuc}{0.68--0.76\xspace}
\newcommand{\zsBestExploitRecall}{51\%\xspace}

\newcommand{\zsBestName}{Gemini 3.1 Flash Lite\xspace}
\newcommand{\zsCut}{13\xspace}
\newcommand{\zsFlagAllFone}{0.64\xspace}

\newcommand{\zsFpr}{0.02--0.06\xspace}

\newcommand{\zsModels}{four\xspace}

\newcommand{\zsRecall}{0.04--0.43\xspace}

\newcommand{\registryCommit}{\code{5e310fc}\xspace}       
\newcommand{\nRegistryFolders}{932\xspace}                
\newcommand{\nScweEntries}{156\xspace}                    
\newcommand{\llamaDefiHacks}{967\xspace}                  
\newcommand{\llamaDefiLoss}{9.29~billion\xspace}          
\newcommand{\pLlamaSeven}{75.4\%\xspace}                  
\newcommand{\eulerLossRegistry}{197~million\xspace}       
\newcommand{\tauUsd}{100\xspace}                          
\newcommand{\nearDays}{30\xspace}                         
\newcommand{\gainMargin}{5\%\xspace}                      
\newcommand{\minNonce}{10\xspace}                         
\newcommand{\rereadDays}{seven\xspace}                       

\newcommand{\hide}[1]{}

\begin{document}

\title{TxBench: Verifying Public Incident Data to Benchmark Smart-Contract Attack Detectors}

\author{\IEEEauthorblockN{Tamer Abdelaziz}
\IEEEauthorblockA{NYU Abu Dhabi \\ tamer.m@nyu.edu}
\and
\IEEEauthorblockN{Karim Ali}
\IEEEauthorblockA{NYU Abu Dhabi \\ karim.ali@nyu.edu}
}

\maketitle

\begin{abstract}
Malicious transactions on smart contracts cause billions of dollars in losses. To evaluate their detectors, most datasets copy the attack transactions that public incident databases cite and pair them with random transactions. However, these databases never check their citations against execution, and random transactions rarely resemble attacks. We present \txbench, a benchmark that verifies both attacks and legitimate transactions against their execution. \txbench labels a cited transaction as an attack only when its execution takes value from others and independent evidence ties it to its incident. Its legitimate transactions perform the same operations as attacks and carry positive evidence of legitimacy. Applying these checks to two public databases, we find that they share only \pBoth of their \nCited~cited transactions and that only \nAccepted are verifiable attacks, worth \$\accValue. A blind manual review confirms \nReviewAConf/\nReviewA sampled attacks and \nReviewBConf/\nReviewB legitimate transactions.
\end{abstract}


\section{Introduction} \label{sec:intro}

\glsresetall

Attacks on smart contracts, the programs that hold \defi assets, cause billions of dollars in losses. DeFiLlama~\cite{defillama2026hacks} records \llamaDefiHacks~hacks worth \$\llamaDefiLoss, each carried out by one or more malicious transactions~\cite{zhou2023sok}. Code analyzers~\cite{abdelaziz2026analyzers,durieux2020smartbugs} help prevent these attacks but flag far more contracts than attackers exploit. For example, attackers exploit only 2\% of the 23,327 contracts that six analyzers flag as vulnerable~\cite{perez2021vulnerable}. Hence, recent detectors inspect each transaction's execution trace to replay it against rules~\cite{zhang2020txspector,torres2021horus}, lift it to \defi actions~\cite{wu2024defiranger,xie2024defort,zhong2025defiscope}, or learn models of it~\cite{wu2025hunting,meng2026houston,abdelaziz2026txlens}.

These transaction-level detectors lack verified labels, because existing benchmarks label code, not executed transactions~\cite{durieux2020smartbugs,zheng2024dappscan,evmbench2026}. Hence, each detector paper builds its own evaluation set. Horus~\cite{torres2021horus} labels the attacks that its own analyses find, making the evaluation circular. Other papers copy attacks from public incident databases~\cite{alhaidari2025onchain,wu2025bridgeguard}. Similar to other mined software data~\cite{croft2022noisy}, these databases may contain label errors that every dataset built on them inherits. A benchmark also needs legitimate transactions to measure false positives, a main barrier to adopting analyzers in practice~\cite{abdelaziz2026analyzers}.
However, datasets leave them unverified~\cite{wu2025bridgeguard}, and random transactions rarely do as much computation as attacks.

The Euler Finance exploit~\cite{eulerhack} shows both problems. BlockSec~\cite{blocksec2026library} lists \eulerTxs~drain transactions and a loss of \$\eulerLossBlockSec, while the EVM Hack Registry~\cite{evmhackregistry} lists 1~transaction and a loss of \$\eulerLossRegistry. Their execution traces show that BlockSec's transactions take \$\eulerValue, while the registry's transaction takes only \$\eulerRegValue. Hence, a dataset built from the registry credits one transaction with all losses and misses the other drain transactions. Moreover, the attacker uses only calls that legitimate users also make (e.g., a donation to the protocol's reserves, a liquidation). Therefore, to show that a detector recognizes the attack rather than its calls, a benchmark needs legitimate transactions that make these calls too.

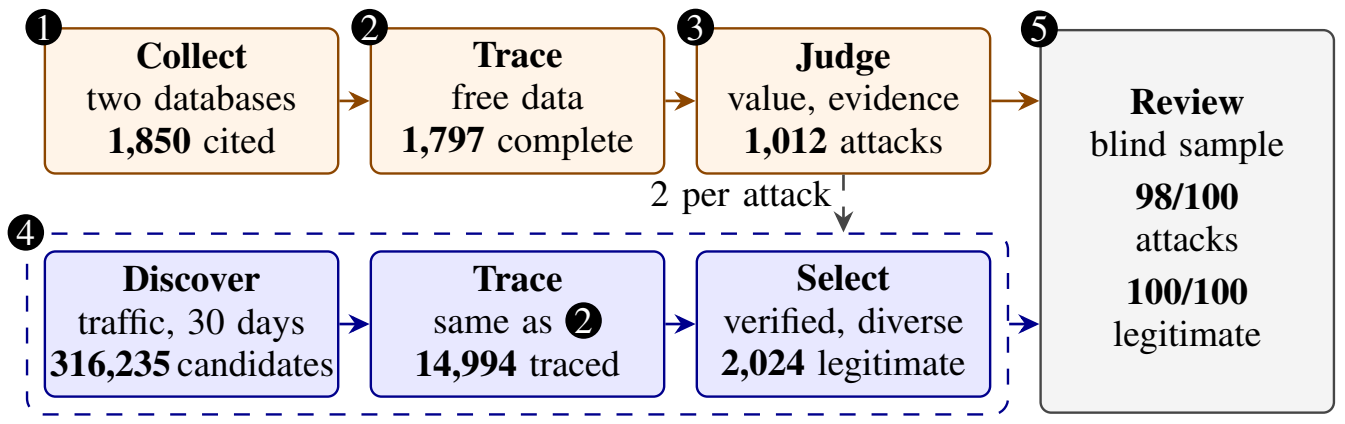
\begin{figure}
\centering
\scriptsize
\begin{tikzpicture}[
  node distance=5mm and 2mm,
  box/.style={draw=#1!55!black, fill=#1!9, rounded corners=2pt, line width=0.45pt, align=center, text width=18.7mm, minimum height=9.6mm, inner sep=1pt},
  step/.style={label={[anchor=center, inner sep=0pt]north west:\circnum{#1}}},
  flow/.style={-{Stealth[length=1.4mm,width=1.3mm]}, line width=0.5pt, draw=#1!55!black},
  flow/.default=black]
\node[box=orange, step=1] (collect) {\textbf{Collect}\\ two databases\\ \mbox{\textbf{\nCited}~cited}};
\node[box=orange, step=2, right=of collect] (trace) {\textbf{Trace}\\ free data\\ \mbox{\textbf{\nFullTrace}~complete}};
\node[box=orange, step=3, right=of trace] (judge) {\textbf{Judge}\\ value, evidence\\ \mbox{\textbf{\nAccepted}~attacks}};
\node[box=blue, below=of collect] (discover) {\textbf{Discover}\\ traffic, 30 days\\ \mbox{\textbf{\nCandidates}\,candidates}};
\node[box=blue, right=of discover] (trace2) {\textbf{Trace}\\ same as~\circnum{2}\\ \mbox{\textbf{\nNegTraced}~traced}};
\node[box=blue, right=of trace2] (select) {\textbf{Select}\\ verified, diverse\\ \mbox{\textbf{\nNeg}~legitimate}};
\node[draw=blue!55!black, dashed, rounded corners=3pt, line width=0.45pt, inner sep=1.1mm, fit=(discover)(select), step=4] (lane) {};
\path (judge.north) -- (lane.south) coordinate[midway] (mid);
\node[box=gray, step=5, anchor=west, minimum height=25.3mm] (review) at ([xshift=2mm]lane.east |- mid) {\textbf{Review}\\ blind sample\\[0.6ex] \textbf{\nReviewAConf/\nReviewA}\\ attacks\\[0.6ex] \textbf{\nReviewBConf/\nReviewB}\\ legitimate};
\draw[flow=orange] (collect) -- (trace);
\draw[flow=orange] (trace) -- (judge);
\draw[flow=orange] (judge.east) -- (review.west |- judge.east);
\draw[flow=blue] (discover) -- (trace2);
\draw[flow=blue] (trace2) -- (select);
\draw[flow=blue] (lane.east |- select.east) -- (review.west |- select.east);
\draw[flow=gray, dashed] (judge) -- node[left, pos=0.4, font=\scriptsize, inner sep=2pt] {2 per attack} (lane.north -| judge);
\end{tikzpicture}
\caption{An overview of the \txbench pipeline. Numbers count transactions.}
\label{fig:pipeline}
\vspace{-2\baselineskip}
\end{figure}

To give transaction-level detectors verified labels, we present \txbench, a benchmark that checks every label against the execution trace of its transaction, covering 7~\evm chains. \Cref{fig:pipeline} shows the pipeline that we use to build \txbench from free public data. We first collect the \nCited~transactions that BlockSec and the EVM Hack Registry cite~(\circnum{1}), and rebuild their traces~(\circnum{2}). We then accept \nAccepted~of them as attacks, requiring that each one takes value from accounts outside the attacker's control and has evidence beyond its citation~(\circnum{3}). For legitimate transactions, we sample each chain's traffic within 30~days of its attacks and keep \nNeg~transactions with evidence of legitimacy~(\circnum{4}). Finally, we manually review a blind sample of both classes~(\circnum{5}). No step uses a detector to decide a label.

Building \txbench reveals how far the two databases are from a ground truth. They share only \nBoth/\nCited cited transactions, and only \pTakeValue of the cited transactions with a complete trace take at least \$\tauUsd from accounts outside the attacker's control. Random transactions also make poor legitimate examples, because only \pBlockGas of them use more than a million gas (the unit of computation that a transaction pays for), compared to \pAccGas of the attacks. We release \txbench, its pipeline, and its caches on Zenodo~\cite{abdelaziz2026txbench}.
\section{From Incident Citations to Accepted Attacks} \label{sec:collection}

\subsection{Sourcing \txbench From Two Public Incident Databases}
To cover more incidents than any single database, we collect the attack transactions that two public databases cite~(\circnum{1}). The first database is BlockSec's Security Incidents Library~\cite{blocksec2026library}, which documents incidents with their transactions, attackers, root causes, and losses. Our snapshot of that library holds \nBlockSecIncidents~incidents, dated from \blocksecFirstDate to \blocksecLastDate. The second database is the EVM Hack Registry~\cite{evmhackregistry}, which extends DeFiHackLabs~\cite{defihacklabs} with \nRegistryFolders~dated incidents from 2017 to 2026 at commit \registryCommit. Each registry incident comes with a report that cites its transactions and parties, and with a Foundry~\cite{foundry} proof of concept that reproduces the attack on a copy of the chain. To combine both databases, we match them on the transaction hash (i.e., the identifier of each transaction). We then keep the 7~chains with more than 10~incidents in their union: Ethereum, \bsc, Arbitrum, Base, Optimism, Polygon, and Avalanche. Every other chain has at most 4~incidents. Attacks on these chains alone account for \pLlamaSeven of the \defi losses that DeFiLlama records~\cite{defillama2026hacks}. This step results in \nCited~cited transactions: \nBoth in both databases, \nBlockSecOnly only in BlockSec, and \nRegistryOnly only in the registry. \Cref{tab:overview} breaks down these counts per chain, along with the attacks and legitimate transactions that later steps keep. Of the cited transactions, \nSucceeded succeed on chain, and \nReverted revert (i.e., they fail and the chain undoes their effects). We drop the remaining \nNonexistent that match no transaction on these chains.

\subsection{Weaknesses in a Shared Taxonomy}
To compare the weakness labels of both databases, and to let researchers measure recall per weakness, we map each incident to \scwe~\cite{owasp2026scwe}, a taxonomy of \nScweEntries~weaknesses~(\circnum{1}). To achieve that, we use \claude to read the root cause of each incident from the registry's report, a post-mortem, the traces, or BlockSec's label, under a fixed codebook that we include in the artifact. All transactions of an incident then share that incident's \scwe entry. Overall, \nScweId/\nCited transactions map to a specific entry, and \nScweNone~transactions belong to attacks that exploit no contract flaw (e.g., key compromises). To validate these readings, one of the authors reads \nScweSheet~incidents blindly under the same codebook, and assigns the same entry as \claude for \nScweRevSet/\nScweSheet ($\kappa = 0.83$).

\subsection{Execution Records from Free Data Sources}
The receipt that chains store for a transaction lists its events but not the internal calls where an attack happens. Therefore, we build an enriched transaction trace that holds its receipt, decoded call and events, and internal call frames~(\circnum{2}). To enable free reproduction, our trace builder reads only public endpoints, free \api tiers and tracers, Sourcify~\cite{sourcify}, and public signature databases. Since \bsc and Arbitrum offer no free tracer, the trace builder replays each of their blocks in a local Anvil fork~\cite{foundry}, keeping the replay only when its status and logs equal those of the chain. Of the \nTraced~transactions on chain, \nFullTrace have the complete call trees that the value oracle needs (\circnum{3}).

\begin{table}
\centering
\caption{\txbench per chain. Value is in \$~million. Incidents have no total, because an incident may span several chains.}
\label{tab:overview}
\begin{tabular}{l r r r r r}
\toprule
& \#Incidents & \#Cited & \#Attacks & Value & \#Legitimate \\
\midrule
Ethereum & 405 & 904 & 544 & 692.4 & 1,088 \\
BSC & 406 & 560 & 309 & 67.8 & 618 \\
Arbitrum & 67 & 183 & 43 & 30.4 & 86 \\
Base & 63 & 107 & 65 & 12.5 & 130 \\
Optimism & 21 & 43 & 25 & 42.3 & 50 \\
Polygon & 28 & 36 & 17 & 10.8 & 34 \\
Avalanche & 14 & 17 & 9 & 12.0 & 18 \\
\midrule
Total & -- & \nCited & \nAccepted & 868.2 & \nNeg \\
\bottomrule
\end{tabular}
\end{table}

\subsection{A Value Oracle Instead of a Detector}
Each database specifies who the attacker and victim of a transaction are and how much its incident loses. Therefore, we judge each transaction by its own execution rather than by these claims~(\circnum{3}). To achieve that, we design a value oracle that computes each address's net change per asset from the trace, and prices it in USD an hour before the block~\cite{defillama2026coins}. This approach prevents the attack from pricing its own gains. The oracle then measures what the attacker side takes from accounts outside it. We call an attack a key compromise when it uses stolen private keys, and a contract exploit otherwise. The attacker side holds the named attackers, the sender of a contract exploit, their created contracts, and wallets that only receive from them. Of the \nJudged~judged transactions, \nTakeValue take at least \$\tauUsd, a floor that filters dust transfers.

We then check each value-taking transaction against 3~database claims: the sender of a contract exploit is a named attacker, a named victim loses value, and the incident's value is within 10$\times$ the reported loss to account for free-text losses priced at different times. Of these transactions, \nFailCheck fail at least one check, leaving \nPassChecks. The loss check fails most often (\nFailLoss), because a database may report all attack losses but cite only some of its transactions. However, the sender and loss checks pass almost by construction. This is because BlockSec usually names the sender itself as the attacker, and a database usually computes its reported loss from the cited transactions. Therefore, we accept a transaction as an attack only if it also has one of the following evidence types: both databases cite it, the registry's proof of concept replays the attack from a block within an hour, the registry names its sender an attacker, a victim that the registry names loses value, or a database names the thief who gains from a key compromise. These checks yield \nAccepted~attacks: \nAccExploit~contract exploits and \nAccKey~key compromises. The other \nSecondary~transactions lack enough evidence and come from BlockSec alone. To measure how far the evidence requirement favors the registry's incidents, we ship them as a secondary set.
\section{Verified Legitimate Transactions} \label{sec:negatives}
To let researchers measure precision as well as recall, we supplement \txbench with legitimate transactions that perform the same kinds of operations as attacks~(\circnum{4}). Such transactions test whether a detector recognizes the attack itself rather than surface features that attacks share with legitimate activity. This section describes how we discover candidates, verify their legitimacy, and sample among the verified ones.

\subsection{Targets and Discovery}
\txbench targets two legitimate transactions per accepted attack on each chain (\nTarget in total), a ratio that keeps the tracing within free \api tiers. Of the accepted attacks, \pAccCreation deploy a contract and then attack from its constructor (i.e., the code that runs once when the contract is created). Hence, each chain's legitimate transactions include two contract deployments per such attack. The remaining legitimate transactions split 50/30/20 over 3~categories: behaviorally hard (e.g., swaps and liquidations), financially hard (i.e., values at or above the 25th percentile of the chain's attack values), and verified ordinary (e.g., transfers and approvals). This split gives the most weight to operations that attacks also use. To collect candidates, we draw seeded random windows within \nearDays~days of each chain's attacks. Most windows are single random blocks. The rest target blocks with liquidations and contract creations, which random blocks rarely hold. These \nWindows~windows hold \nCandidates~on-chain candidate transactions. We then trace \nNegTraced of them with the same trace builder as the attacks~(\circnum{2}), decoding all but one.

\subsection{Positive Evidence of Legitimacy}
Calling well-known protocols does not make a transaction legitimate, because attackers also use them to swap stolen funds~\cite{adams2021uniswap} or borrow funds~\cite{frangella2022aave}. Absence from the incident databases does not make it legitimate either, because the databases may miss some incidents. Therefore, we require positive evidence of legitimacy, and sort candidates into three tiers by that evidence. Only tier-1 candidates enter \txbench. A candidate is tier~1 only when (1)~it succeeds with a complete call tree; (2)~no third party loses unexplained value beyond the larger of 1\% of the transaction's volume and \$100; (3)~every venue that emits a market event in it is one of \nRegistryRows~deployments of \nRegistryProtocols~protocols that we check on chain, or a pool/market that such a deployment lists; (4)~its sender gains at most the larger of \gainMargin of the volume and \$100, or 15\% of the volume for a liquidation, margins that legitimate arbitrage and liquidations rarely exceed; and (5)~it involves no cited transaction, no named or oracle-found attacker, and no incident's main victim from one day before to 30~days after the incident. A candidate that lacks evidence is tier~2, and we ship these separately. A candidate that contradicts a rule is tier~3. Of the \nLabeled~traced candidates, \nTierOne are tier~1.

\subsection{Sampling under Diversity Caps}
We sample only tier-1 candidates, in cells of chain, category, and year. We draw each behavioral and financial cell evenly across the quartiles of the chain's attack gas use. We then cap the legitimate transactions at 2~per sender, 20~per called contract, and 3~per block, with no protocol $>$ 25\% of a chain and category. Since \pAccFresh of attacks come from fresh senders (i.e., senders with $<$ \minNonce~prior transactions), each cell also takes fresh senders in the same share. This process yields \nNeg~legitimate transactions: \nNegDeploy~deployments, \nNegBehavioral~behavioral, \nNegFinancial~financial, and \nNegOrdinary~ordinary. These transactions resemble attacks more than random traffic does. Of them, \pNegGas use over a million gas (5.9\% without deployments), compared to \pBlockGas of random transactions and \pAccGas of attacks. Hence, gas use alone nearly identifies attacks in random traffic, while \txbench forces a detector to look beyond.
\section{Measuring Label Quality before Trusting It} \label{sec:review}
\txbench assigns every label by rule. Hence, a wrong label distorts every metric that a detector reaches on \txbench. To estimate how often the rules err, one of the authors manually reviews a sample of attacks and legitimate transactions~(\circnum{5}). Before drawing the sample, we fix the decision rules for the review by recording their cryptographic hash, preventing any change after the reviewer sees the data. These rules ask one question of each transaction: \emph{does it take, or directly enable taking, assets or control from a victim without consent, through a contract flaw, a stolen key or signature, or deception?} The reviewer answers \code{attack\_yes}, \code{attack\_no}, or \code{unsure}.

The reviewed sample consists of \nReviewA~accepted attacks and \nReviewB~legitimate transactions, in \nReviewStrata~strata by chain and period, interleaved without labels, keeping the review blind. To measure consistency, the reviewer reads a seeded fifth of the transactions again at least \rereadDays~days later, without seeing the first decision. The two readings agree on \pAgree of the \nOverlap re-read transactions ($\kappa = \kappaReview$, \ci of \kappaCI). Adjudication settles the \nDisagree disagreements. The review confirms \nReviewAConf/\nReviewA attacks and \nReviewBConf/\nReviewB legitimate transactions, counting \code{unsure} as not confirmed. Weighting each stratum by its share of \txbench, \pReviewA of the accepted attacks carry the right label, with a \ci of \ciReviewA. For legitimate transactions, the estimate is \pReviewB, with a \ci of \ciReviewB. The review rejects \nReviewARejected~attacks, a key-compromise transfer and a deployment that performs a transfer. Both retain their rule-assigned labels, keeping the estimate valid for every label in \txbench. The artifact ships every review decision, enabling users to drop such transactions.
\section{Zero-Shot LLMs Miss Most Attacks} \label{sec:baselines}
To show how challenging \txbench is for detectors that need no training data, we ask \zsModels~zero-shot \llms about the reviewed transactions. We then score their answers against the review's final decisions. Each prompt holds a transaction's trace file as untruncated JSON and asks whether the transaction is an attack. To keep an \llm from recognizing a known incident by its identity, we replace every address with a consistent placeholder, and remove the hashes, block, and entity labels. The test holds the \nZsAsked~reviewed transactions whose prompts fit every \llm context window: \nZsAttacks~attacks (\nZsExploits~contract exploits and \nZsKey~key compromises) and \nZsNegatives~legitimate transactions. The rest~(\nZsNotAsked) are contract exploits with thousands of internal calls. We run each \llm once at temperature zero. Since the answers may still vary between runs, we release them all with \txbench. The answers that run out of output tokens~(\zsCut) count as misses.

\Cref{tab:zeroshot} shows that zero-shot \llms flag attacks precisely but miss most of them. Since \pZsAlways of the test transactions are attacks, flagging every transaction (Flag All) already reaches an F1 of \zsFlagAllFone, above every \llm. Therefore, we compare the \llms by AUC and recall. All \llms rank attacks above legitimate transactions better than chance (AUC \zsAuc, against 0.50 for Flag All). However, they miss most attacks (recall \zsRecall), while rarely flagging a legitimate transaction (false-positive rate \zsFpr). As expected, no \llm catches a key compromise, because most are plain transfers that no trace can reveal. Even the best \llm, \zsBestName, finds only \zsBestExploitRecall of the contract exploits. Hence, \txbench leaves most attacks for future detectors to find.
\section{Storage, Use, and Research Directions} \label{sec:dataset}
\txbench consists of 3~files: accepted attacks, legitimate transactions, and a combination of both in \code{combined.csv}. Each row of \code{combined.csv} carries a transaction's label, category, and 5~train/test splits, where no group appears on both sides of a split. A data dictionary defines every column. To keep file paths from revealing the class, all enriched traces sit in one flat folder, named by transaction hash. Contract exploits form the primary evaluation, and key compromises form a separate track. The script \code{evaluate.py} scores a detector's predictions per track and split, weighting incidents equally to keep large incidents from dominating the score. It also reweights precision for the real-traffic share of attacks, which \txbench's 2~legitimate transactions per attack do not reflect.

All outputs, including the traces and the caches, take \sizeGB~GB. From these caches, \code{run\_pipeline.py --offline} rebuilds every output in standard-library Python, showing that the pipeline is deterministic. An online rerun requeries the incident databases and chains, and needs only free NodeReal and Alchemy \api keys and Foundry's Anvil (see the README). Beyond evaluation, \txbench supports four lines of research: precision against legitimate transactions with the same operations as attacks, generalization across time and chains, recall per weakness and weighted by USD, and the reliability of public incident data. For the last one, the oracle checks record where each database contradicts the chain. As a new resource, \txbench has no usage history.
\section{Limitations and Improvements} \label{sec:limitations}

\paragraph{Internal Validity} \textbf{I1.} One author reviews every sampled transaction, and the agreement between the two readings measures only that reviewer's consistency. With \nReviewA~transactions per class, the attacks' interval also stays wide (\ciReviewA), although the hashed rules and released decisions let others repeat the review. \textbf{I2.} A list of \nRegistryRows~protocol deployments decides which venues \txbench recognizes. No tier-1 legitimate transaction trades through an unrecognized venue, while \nAccUnvouched~accepted attacks (\pAccUnvouched) do, letting a detector learn the list instead of the attacks. The tier-2 transactions include such venues, enabling users to test for this shortcut.

\paragraph{External Validity} \txbench holds only the attacks that the two incident databases cite. BlockSec's library starts in 2023, leaving only \nAccPreTwentyThree~accepted attacks from earlier years.

\paragraph{Construct Validity} \textbf{C1.} Of the \nAccKey~key compromises, \nKeyPlain are plain transfers that no trace can reveal. \topIncidentName alone contributes \nTopIncident~attacks (\pTopIncident). Hence, \txbench evaluates key compromises separately and weights incidents equally. \textbf{C2.} The \nNegDeploy~deployments are easy to detect by counting calls, because \pNegDeployNoCall of them make no calls (AUC \aucCallsDep). Hence, users should report results with and without them.

\paragraph{Improvements} \txbench can improve by adding legitimate transactions that pay their sender, multi-transaction attacks, more protocols, and harder deployments.
\section{Related Work} \label{sec:related}

\txbench builds on work in three areas. First, benchmarks for smart-contract security, such as SmartBugs~\cite{durieux2020smartbugs}, DAppSCAN~\cite{zheng2024dappscan}, and EVMbench~\cite{evmbench2026}, label code rather than executed transactions. Unlike \txbench, they cannot measure whether a detector flags an attack in the transactions that carry it out. Second, incident databases~\cite{blocksec2026library,defihacklabs,evmhackregistry} cite the transactions behind each reported attack without checking them against execution. As a result, transaction-level detectors~\cite{zhang2020txspector,torres2021horus,wu2024defiranger,xie2024defort,zhong2025defiscope,wu2025hunting,meng2026houston,abdelaziz2026txlens} each assemble their own dataset, labeled by the detector itself, copied from such databases, or paired with random legitimate transactions. \txbench gives these detectors one shared dataset that verifies every cited transaction against its execution and pairs the attacks with evidence-backed legitimate transactions. Third, prior research shows that mined labels are often wrong, for both defect reports~\cite{DBLP:conf/icse/HerzigJZ13} and security defects~\cite{croft2022noisy}. \txbench brings the same scrutiny to incident data, measuring the error that remains.

\begin{table}
\centering
\caption{Zero-shot \llms on the \nZsAsked~reviewed transactions. \emph{Flag All} labels every transaction as an attack.}
\label{tab:zeroshot}
\begin{tabular}{l r r r r}
\toprule
\llm & Precision & Recall & F1 & AUC \\
\midrule
Gemini 3.1 Flash Lite & 0.93 & 0.43 & 0.59 & 0.76 \\
GLM 5.3 Flash & 0.81 & 0.29 & 0.42 & 0.72 \\
DeepSeek V4 Flash & 0.88 & 0.24 & 0.38 & 0.74 \\
Qwen 3.7 Flash & 0.67 & 0.04 & 0.08 & 0.68 \\
\midrule
Flag All & 0.47 & 1.00 & 0.64 & 0.50 \\
\bottomrule
\end{tabular}
\end{table}        
\section{Conclusion} \label{sec:conclusion}
\txbench is a benchmark of \nAccepted~attacks and \nNeg~evidence-backed legitimate transactions on 7~\evm chains, where execution and evidence decide every label. Building it shows that the two incident databases share only \pBoth of their \nCited~cited transactions. Hence, datasets that copy these citations inherit their errors. Since \txbench builds on free public data, it may grow with each new incident. We archive the dataset, pipeline, tests, and review protocol on Zenodo~\cite{abdelaziz2026txbench}, with the code under MIT and our annotations under CC~BY~4.0. Third-party material keeps its terms.

\bibliographystyle{IEEEtranN}
\bibliography{references}

\end{document}